# Problematic zeolite records and bibliographic reference link for normal zeolite records in the Inorganic Crystal Structure Database

S. Yang, M. Lach-hab, I. I. Vaisman, X. Li and E. Blaisten-Barojas

*Computational Materials Science Center, George Mason University, MSN 6A2, Fairfax, Virginia 22030*

(August 27, 2009)

## Abstract

In this work we provide a list of 200 crystal entries in the Inorganic Crystal Structure Database (ICSD) that have errors, incomplete information, or are given a generic name of zeolite when they may belong to a different family of crystals. We also provide a table containing the ICSD code number of 1473 zeolite entries and the correspondence to the original bibliographic reference.

## 1 Database entries with problems

The Inorganic Crystal Structure Database (ICSD) [1] has the most complete crystal structure collection gathered up to date. In analyzing entries corresponding to the family of zeolites our group discovered that there are 200 records corresponding to crystals containing incomplete or inaccurate information. Twenty of these crystal entries are listed in the database with non-standard space group notation (with centered enlarged unit cells) as listed in Table 1. Additionally, the ICSD contains 24 entries with minor errors, which we believe can be corrected as indicated in Table 2. However, our proposed corrections require expert review from the organizations maintaining and producing the ICSD.

The ICSD has 98 crystals that are referred to with the generic name "zeolite." These crystals should be classified or named based on their crystallographic information and further identification research is needed on them in order to be useful to the community of scientists working with zeolites. The list is included in Table 3. Crystals are recognized to belong to the family of zeolites if a framework type code [2] can be assigned. In our studies [3, 4, 5, 6, 7, 8] we could not assign a framework type to 58 crystals that are named "zeolites" in the ICSD records. These entries are given in Table 4.

In previous machine learning studies [3, 4, 5, 6, 7, 8] a dataset of 1473 zeolite entries from the ICSD were used, and in Ref. 4 the framework type was assigned and confirmed for 1436 of them. In these works the ICSD number was used to identify each zeolite structure. Table 5 provides a correspondence between the ICSD number and the bibliographic reference that originated the database entry. This table should be useful to researchers that do not have access to the ICSD.

## 2 Conclusions

Findings contained in Tables 1-4 have been communicated on August 26, 2008 to the producers of the ICSD at the National Institute of Standards and Technology for inclusion in the database. Table 5 provides a strict correspondence between the original bibliographic reference and the code number assigned to that entry by the ICSD producers.

**Acknowledgments**

This work was supported under the National Science Foundation grant CHE-0626111. Authors gratefully acknowledge the Standard Reference Data Program of the National Institute of Standards and Technology for making available the data set from the ICSD.

**Table 1.** Twenty crystals reported in the ICSD with a non-standard space group notation.

| Space group | ICSD # |
|---|---|
| C 1 | 82770 |
| C -1 | 88921, 151181 |
| F 1 1 2 | 83008, 83009, 83010, 83011, 83012, 31249, 81104 |
| C 1 1 21 | 73364, 81392 |
| F 1 d 1 | 172122, 54890, 30804, 50060, 59416 |
| B 1 21/d 1 | 98118 |
| F 1 2/m 1 | 83467, 83468 |

**Table 2.** List of 24 entries in the ICSD with minor problems and the suggested correction.

| ICSD # | Min Name | Corrected |
|---|---|---|
| 40644 | Zeolite ANA Ga | Ga/Si have different coordinates |
| 60891 | Zeolite A | mineral name should be Zeolite ABW |
| 60892 | Zeolite A | mineral name should be Zeolite ABW |
| 68257 | Zeolite A | mineral name should be Zeolite ABW |
| 60890 | Zeolite Li-A | mineral name should be Zeolite Li-ABW |
| 40128 | Zeolite Li-A | mineral name should be Zeolite Li-ABW |
| 68101 | Zeolite Li-A | mineral name should be Zeolite Li-ABW |
| 89954 | Zeolite Li-A | mineral name should be Zeolite Li-ABW |
| 40941 | Zeolite | mineral name should be Zeolite ABW |
| 89055 | Brewsterite | short T-T distance; corrected with Monte Carlo simulation |
| 93955 | Brewsterite | short T-T distance; corrected with Monte Carlo simulation |
| 91699 | Brewsterite | short T-T distance; corrected with Monte Carlo simulation |
| 91700 | Brewsterite | short T-T distance; corrected with Monte Carlo simulation |
| 201472 | Zeolite NaY | inconsistency in *Structured*, *Sum*, and *Analytical*: Na57 |
| 97918 | Ferierite | mineral name should be Ferrierite |
| 97919 | Ferierite | mineral name should be Ferrierite |
| 97920 | Ferierite | mineral name should be Ferrierite |
| 172576 | Zeolite TNU-7 | coordinates wrong; corrected based on the original paper |
| 172072 | Estilbite | mineral name should be Epistilbite |
| 172073 | Estilbite | mineral name should be Epistilbite |
| 172074 | Estilbite | mineral name should be Epistilbite |
| 172075 | Estilbite | mineral name should be Epistilbite |
| 31383 | Dachiardite | short T-T distance; corrected with Monte Carlo simulation |
| 85550 | Tschoertnerite | mineral name should be Tschörtnerite |

**Table 3.** List of 98 entries in the ICSD referred to with the generic denomination "zeolite."

| Min Name | ICSD # | | | | | | | | | |
|---|---|---|---|---|---|---|---|---|---|---|
| Zeolite | 9463 | 9624 | 9625 | 9626 | 20645 | 30919 | 40272 | 66154 | 68920 | 72818 |
| | 201183 | 407793 | 40930 | 40941 | 88456 | 170475 | 170476 | 170477 | 170478 | 170479 |
| | 170480 | 170481 | 170482 | 170483 | 170484 | 170485 | 170486 | 170487 | 170488 | 170489 |
| | 170490 | 170492 | 170493 | 170494 | 170495 | 170496 | 170497 | 170498 | 170499 | 170500 |
| | 170501 | 170502 | 170503 | 170504 | 170505 | 170506 | 170507 | 170508 | 170509 | 170510 |
| | 170511 | 170512 | 170513 | 170514 | 170515 | 170516 | 170517 | 170518 | 170519 | 170520 |
| | 170521 | 170522 | 170523 | 170524 | 170525 | 170526 | 170527 | 170528 | 170529 | 170530 |
| | 170531 | 170532 | 170533 | 170534 | 170535 | 170536 | 170537 | 170538 | 170539 | 170540 |
| | 170541 | 170542 | 170543 | 170544 | 170545 | 170546 | 170547 | 170548 | 170549 | 170550 |
| | 170551 | 170552 | 170553 | 170554 | 170559 | 170561 | 170909 | 150188 | | |

**Table 4.** List of 58 crystalline structures with identification "zeolite" distributed over 46 mineral names for which the framework type could not be determined.

| Min Name | Freq. | ICSD # |
|---|---|---|
| Magadiite | 1 | 56290 |
| Nu-6(1) | 1 | 413852 |
| RUB-15 | 1 | 56776 |
| RUB-18 | 1 | 57135 |
| Zeolite ERS-12 | 2 | 151479, 151480 |
| UCR-21GaGeS-AEM sulfide zeolite | 1 | 281755 |
| UCR-21GaGeS-AEP sulfide zeolite | 1 | 281756 |
| UCR-21GaSnS-APP sulfide zeolite | 1 | 281761 |
| UCR-21GaSnSe-TAEA selenide zeolite | 1 | 281760 |
| UCR-21GaSnS-TAEA sulfide zeolite | 1 | 281739 |
| UCR-21GeGaS-APM sulfide zeolite | 1 | 281736 |
| UCR-21GeGaS-APO sulfide zeolite | 1 | 281737 |
| UCR-21GeGaS-HMI sulfide zeolite | 1 | 281738 |
| UCR-21InGeS-AEM sulfide zeolite | 2 | 281740, 281741 |
| UCR-21InGeS-PEHA sulfide zeolite | 1 | 281762 |
| UCR-22GaGeS-AEP sulfide zeolite | 1 | 281742 |
| UCR-22GaSnS-AEP sulfide zeolite | 1 | 281743 |
| UCR-22GaSnSe-TOTDA selenide zeolite | 1 | 281763 |
| UCR-22InGeS-AEP sulfide zeolite | 1 | 281744 |
| UCR-22InGeS-TAEA sulfide zeolite | 1 | 281745 |
| UCR-22InGeS-TETA sulfide zeolite | 1 | 281747 |
| UCR-23GaGeS-AEM sulfide zeolite | 1 | 281748 |
| UCR-23GaSnS-AEM sulfide zeolite | 1 | 281764 |
| UCR-23InGeS-AEM sulfide zeolite | 1 | 281749 |
| Unnamed_Zeolite | 4 | 77238, 77239, 77240, 90652[a] |
| Zemannite | 5 | 33638, 79848, 79849, 80439, 43014 |
| Zeolite ETS-10 | 1 | 77940 |
| Zeolite JDF-20 | 1 | 77628 |
| Zeolite MCM-47 | 1 | 89982 |
| Zeolite MCM-70 | 2 | 153477, 153478 |
| Zeolite MU-15 | 1 | 90628 |
| Zeolite RUB-23 | 1 | 91666 |
| Zeolite RUB-29 | 1 | 90147 |
| Zeolite RUB-31 | 2 | 95378, 95379 |
| Zeolite ULM-10 | 1 | 77672 |
| Zeolite ULM-18 | 1 | 86772 |
| Zeolite ULM-3 | 1 | 77673 |
| Zeolite ULM-4 | 1 | 77926 |
| Zeolite ULM-5 | 1 | 77658 |
| Zeolite ULM-7V | 1 | 77657 |
| Zeolite UND-1 | 1 | 85118 |
| Zeolite VAs-Dab | 1 | 77823 |
| Zeolite VAs-TMA | 1 | 77824 |
| Zeolite ZSM-48 | 2 | 62584, 62585 |
| Zeolite ZZ1 | 1 | 41046 |
| Zirconolite 3T | 1 | 97419 |

[a] 90652 is NAT type.

**Table 5.** The ICSD number and corresponding bibliographic reference for 1473 entries in the database.

| ICSD | Bibliographic reference |
| --- | --- |
| 323 | Galli, E., Acta Crystallographica B, v32, 1623-1627 (1976) |
| 331 | Tambuyzer, E., et al., Acta Crystallographica B, v32, 1714-1719 (1976) |
| 1136 | Schlenker, J.L., et al., Acta Crystallographica B, v33, 2907-2910 (1977) |
| 1234 | Schlenker, J.L., et al., Acta Crystallographica B, v33, 3265-3268 (1977) |
| 2317 | Rinaldi, R., et al., Acta Crystallographica B, v30, 2426-2433 (1974) |
| 2318 | Rinaldi, R., et al., Acta Crystallographica B, v30, 2426-2433 (1974) |
| 2747 | Gard, J.A., et al., Acta Crystallographica B, v28, 825-834 (1972) |
| 4349 | Alberti, A., Tschermaks Mineralogische und Petrographische Mitteilungen, v22, 25-37 (1975) |
| 4361 | Merlino, S., et al., Tschermaks Mineralogische und Petrographische Mitteilungen, v22, 117-129 (1975) |
| 4387 | Leung, P.C.W., et al., Journal of Physical Chemistry, v79, 2157-2162 (1975) |
| 4388 | Vance, T.B.jr., et al., Journal of Physical Chemistry, v79, 2163-2167 (1975) |
| 4389 | Riley, P.E., et al., Inorganic Chemistry, v14, 714-721 (1975) |
| 4390 | Riley, P.E., et al., Journal of the American Chemical Society, v97, 537-542 (1975) |
| 4391 | Riley, P.E., et al., Journal of Physical Chemistry, v79, 1594-1601 (1975) |
| 4392 | Maxwell, I.E., et al., Journal of Physical Chemistry, v79, 1874-1879 (1975) |
| 4393 | Mortier, W.J., et al., Materials Research Bulletin, v10, 1037-1046 (1975) |
| 4394 | Mortier, W.J., et al., Materials Research Bulletin, v10, 1319-1326 (1975) |
| 4395 | Galli, E., et al., Bulletin de la Societe Francaise de Mineralogie et de Cristallographie, v98, 11-18 (1975) |
| 4396 | Galli, E., et al., Bulletin de la Societe Francaise de Mineralogie et de Cristallographie, v98, 331-340 (1975) |
| 6250 | Kocman, V., et al., American Mineralogist, v59, 71-78 (1974) |
| 6258 | Galli, E., Crystal Structure Communications, v3, 339-344 (1974) |
| 6269 | Kerr, I.S., Zeitschrift fuer Kristallographie, Kristallgeometrie, Kristallphysik, Kristallchemie, v139, 186-195 (1974) |
| 6270 | Yanagida, R.Y., et al., Inorganic Chemistry, v13, 723-727 (1974) |
| 6271 | Riley, P.E., et al., Inorganic Chemistry, v13, 1355-1360 (1974) |
| 6272 | Baerlocher, C., et al., Zeitschrift fuer Kristallographie, Kristallgeometrie, Kristallphysik, Kristallchemie, v140, 10-26 (1974) |
| 6273 | de Boer, J.J., et al., Journal of Physical Chemistry, v78, 2395-2399 (1974) |
| 6313 | Baerlocher, C., et al., Helvetica Chimica Acta, v53, 1285-1293 (1970) |
| 6314 | Bartl, H., Neues Jahrbuch fuer Mineralogie. Monatshefte, v1970, 298-310 (1970) |
| 6315 | Olson, D.H., Journal of Physical Chemistry, v74, 2758-2763 (1970) |
| 8186 | Faelth, L., et al., Acta Crystallographica B, v35, 1877-1880 (1979) |
| 8253 | Alberti, A., et al., Acta Crystallographica B, v35, 2866-2869 (1979) |
| 8287 | Galli, E., Rendiconti della Societa Italiana di Mineralogia e Petrologia, v31, 599-612 (1975) |
| 8291 | Mortier, W.J., et al., Zeitschrift fuer Kristallographie, Kristallgeometrie, Kristallphysik, Kristallchemie, v144, 32-41 (1976) |
| 8292 | Mortier, W.J., et al., Zeitschrift fuer Kristallographie, Kristallgeometrie, Kristallphysik, Kristallchemie, v143, 319-332 (1976) |
| 9007 | Maxwell, I.E., et al., Journal of Physical Chemistry, v79, 1874-1879 (1975) |
| 9094 | Galli, E., Acta Crystallographica B, v27, 833-841 (1971) |
| 9262 | Alberti, A., Tschermaks Mineralogische und Petrographische Mitteilungen, v19, 173-184 (1973) |
| 9326 | Smith, J.V., et al., Zeitschrift fuer Kristallographie, Kristallgeometrie, Kristallphysik, Kristallchemie, v126, 135-142 (1968) |

| ICSD | Bibliographic reference |
|---|---|
| 28038 | Barrer, R.M., et al., Zeitschrift fuer Kristallographie, Kristallgeometrie, Kristallphysik, Kristallchemie, v142, 82-98 (1975) |
| 28079 | Scherzer, J., et al., Journal of Physical Chemistry, v79, 1194-1199 (1975) |
| 28080 | Scherzer, J., et al., Journal of Physical Chemistry, v79, 1194-1199 (1975) |
| 28249 | Olson, D.H., et al., Journal of Colloid and Interface Science, v28, 305-314 (1968) |
| 28250 | Olson, D.H., et al., Journal of Colloid and Interface Science, v28, 305-314 (1968) |
| 28251 | Olson, D.H., et al., Journal of Colloid and Interface Science, v28, 305-314 (1968) |
| 28252 | Olson, D.H., et al., Journal of Colloid and Interface Science, v28, 305-314 (1968) |
| 28270 | Galli, E., et al., Mineralogica et Petrographica Acta, v12, 1-10 (1966) |
| 28316 | Gottardi, G., et al., Zeitschrift fuer Kristallographie, Kristallgeometrie, Kristallphysik, Kristallchemie, v119, 53-64 (1963) |
| 28369 | Peacor, D.R., American Mineralogist, v58, 676-680 (1973) |
| 28504 | Olson, D.H., Journal of Physical Chemistry, v72, 4366-4373 (1968) |
| 28508 | Olson, D.H., et al., Nature, v215, 270-271 (1967) |
| 29070 | Dent, L.S., et al., Nature, v181, 1794-1796 (1958) |
| 29505 | Kvick, A., et al., Zeitschrift fuer Kristallographie, v174, 265-281 (1986) |
| 29516 | Belitsky, I.A., et al., Neues Jahrbuch fuer Mineralogie. Monatshefte, v1986, 541-551 (1986) |
| 29517 | Belitsky, I.A., et al., Neues Jahrbuch fuer Mineralogie. Monatshefte, v1986, 541-551 (1986) |
| 29522 | Mazzi, F., et al., Neues Jahrbuch fuer Mineralogie. Monatshefte, v1986, 219-228 (1986) |
| 29539 | Alberti, A., et al., Zeitschrift fuer Kristallographie, v173, 257-265 (1985) |
| 29551 | Bissert, G., et al., Neues Jahrbuch fuer Mineralogie. Monatshefte, v1986, 241-252 (1986) |
| 30003 | Kokotailo, G.T., et al., Nature, v275, 119-1120 (1978) |
| 30018 | Pluth, J.J., et al., Journal of Physical Chemistry, v83, 741-749 (1979) |
| 30120 | Barrer, R.M., et al., Zeitschrift fuer Kristallographie, Kristallgeometrie, Kristallphysik, Kristallchemie, v128, 352-370 (1969) |
| 30278 | Taylor, W.H., et al., Zeitschrift fuer Kristallographie, Kristallgeometrie, Kristallphysik, Kristallchemie, v84, 373-398 (1933) |
| 30279 | Taylor, W.H., et al., Zeitschrift fuer Kristallographie, Kristallgeometrie, Kristallphysik, Kristallchemie, v84, 373-398 (1933) |
| 30597 | Kim, Y., et al., Journal of the American Chemical Society, v100, 175-180 (1978) |
| 30598 | McCusker, L.B., et al., Journal of the American Chemical Society, v101, 5235-5239 (1979) |
| 30599 | McCusker, L.B., et al., Journal of the American Chemical Society, v101, 5235-5239 (1979) |
| 30689 | Mazzi, F., et al., Neues Jahrbuch fuer Mineralogie. Monatshefte, v1984, 373-382 (1984) |
| 30690 | Mazzi, F., et al., Neues Jahrbuch fuer Mineralogie. Monatshefte, v1984, 373-382 (1984) |
| 30740 | Gellens, L.R., et al., Journal of the American Chemical Society, v105, 51-55 (1983) |
| 30791 | Gies, H., Zeitschrift fuer Kristallographie, v164, 247-257 (1983) |
| 30812 | Annehed, H., et al., Zeitschrift fuer Kristallographie, v166, 301-306 (1984) |
| 30873 | Tillmanns, E., et al., Neues Jahrbuch fuer Mineralogie. Monatshefte, v1984, 547-558 (1984) |
| 30918 | Engel, N., et al., Zeitschrift fuer Kristallographie, v169, 165-175 (1984) |
| 30929 | Gramlich-Meier, R., et al., Zeitschrift fuer Kristallographie, v169, 201-210 (1984) |
| 30967 | Kvick, A., et al., Zeitschrift fuer Kristallographie, v171, 141-154 (1985) |
| 30975 | McCusker, L.B., et al., Zeitschrift fuer Kristallographie, v171, 281-289 (1985) |
| 31016 | Galli, E., International Conference on Zeolites: Proceedings 1980, v1980, 205-213 (1980) |
| 31043 | Pearce, J.R., et al., International Conference on Zeolites: Proceedings 1980, v1980, 261-268 (1980) |

| ICSD | Bibliographic reference |
| --- | --- |
| 34179 | Bresciani-Pahor, N., et al., Journal of the Chemical Society. Dalton Transactions, Inorganic Chemistry, v1980, 1511-1514 (1980) |
| 34180 | Bresciani-Pahor, N., et al., Journal of the Chemical Society. Dalton Transactions, Inorganic Chemistry, v1980, 1511-1514 (1980) |
| 34277 | Marti, J., et al., Journal of Colloid and Interface Science, v60, 82-86 (1977) |
| 34296 | Raghavan, N.V., et al., Journal of Physical Chemistry, v80, 2133-2137 (1976) |
| 34297 | Mortier, W.J., et al., Journal of Physical Chemistry, v83, 2263-2266 (1979) |
| 34298 | Mortier, W.J., et al., Journal of Physical Chemistry, v83, 2263-2266 (1979) |
| 34299 | Olsen, D.H., et al., Journal of Physical Chemistry, v85, 2238-2243 (1981) |
| 34329 | Bennett, J.M., et al., Materials Research Bulletin, v3, 865-876 (1968) |
| 34330 | Bennett, J.M., et al., Materials Research Bulletin, v3, 933-940 (1968) |
| 34331 | Bennett, J.M., et al., Materials Research Bulletin, v4, 343-347 (1969) |
| 34364 | Bennett, J.M., et al., Materials Research Bulletin, v4, 77-86 (1969) |
| 34365 | Bennett, J.M., et al., Materials Research Bulletin, v3, 865-876 (1968) |
| 34370 | Flanigan, E.M., et al., Nature, v271, 512-516 (1978) |
| 34396 | Bergerhoff, G., et al., Neues Jahrbuch fuer Mineralogie. Monatshefte, v1958, 193-200 (1958) |
| 34452 | Gordon, E.K., et al., Science, v154, 1004-1007 (1966) |
| 34654 | Smith, J.V., et al., Acta Crystallographica, v17, 374-384 (1964) |
| 34655 | Smith, J.V., et al., Acta Crystallographica, v17, 374-384 (1964) |
| 34807 | Baur, W.H., American Mineralogist, v49, 697-704 (1964) |
| 34890 | Meier, W.M., Zeitschrift fuer Kristallographie, Kristallgeometrie, Kristallphysik, Kristallchemie, v113, 430-444 (1960) |
| 34891 | Meier, W.M., Zeitschrift fuer Kristallographie, Kristallgeometrie, Kristallphysik, Kristallchemie, v115, 439-450 (1961) |
| 34937 | Thoeni, W., Zeitschrift fuer Kristallographie, Kristallgeometrie, Kristallphysik, Kristallchemie, v142, 142-160 (1975) |
| 34938 | Thoeni, W., Zeitschrift fuer Kristallographie, Kristallgeometrie, Kristallphysik, Kristallchemie, v142, 142-160 (1975) |
| 34939 | Thoeni, W., Zeitschrift fuer Kristallographie, Kristallgeometrie, Kristallphysik, Kristallchemie, v142, 142-160 (1975) |
| 35078 | Gottardi, G., et al., Zeitschrift fuer Kristallographie, Kristallgeometrie, Kristallphysik, Kristallchemie, v119, 53-64 (1963) |
| 35082 | Vaughan, P.A., Acta Crystallographica, v21, 983-990 (1966) |
| 35119 | Adams, J.M., et al., Journal of Solid State Chemistry, v44, 245-253 (1982) |
| 35120 | Adams, J.M., et al., Journal of Solid State Chemistry, v44, 245-253 (1982) |
| 35320 | Fang, J.H., Thesis, Univ. Pennsylvania, v1961, n9, 1-1 (1961) |
| 35572 | Parise, J.B., et al., Materials Research Bulletin, v18, 841-852 (1983) |
| 35573 | Parise, J.B., et al., Materials Research Bulletin, v18, 841-852 (1983) |
| 35574 | Parise, J.B., et al., Materials Research Bulletin, v18, 841-852 (1983) |
| 35680 | Pluth, J.J., et al., Journal of the American Chemical Society, v105, 1192-1195 (1983) |
| 35681 | Pluth, J.J., et al., Journal of the American Chemical Society, v105, 1192-1195 (1983) |
| 35691 | Pluth, J.J., et al., Journal of the American Chemical Society, v105, 2621-2624 (1983) |
| 35692 | Pluth, J.J., et al., Journal of the American Chemical Society, v105, 2621-2624 (1983) |
| 35693 | Pluth, J.J., et al., Journal of the American Chemical Society, v105, 2621-2624 (1983) |
| 35694 | Pluth, J.J., et al., Journal of the American Chemical Society, v105, 1611-1614 (1983) |

| ICSD | Bibliographic reference |
| --- | --- |
| 35695 | Pluth, J.J., et al., Journal of the American Chemical Society, v105, 2621-2624 (1983) |
| 36193 | Vaughan, P.A., Acta Crystallographica, v21, 983-990 (1966) |
| 36206 | Simpson, H.D., et al., Journal of the American Chemical Society, v91, 6229-6232 (1969) |
| 36323 | Dent, L.S., et al., American Crystallographic Association: Program and Abstracts, v1958, 48-48 (1958) |
| 37061 | Galli, E., et al., Acta Crystallographica B, v39, 189-197 (1983) |
| 37062 | Galli, E., et al., Acta Crystallographica B, v39, 189-197 (1983) |
| 37063 | Galli, E., et al., Acta Crystallographica B, v39, 189-197 (1983) |
| 37064 | Galli, E., et al., Acta Crystallographica B, v39, 189-197 (1983) |
| 37148 | Wyart, J., Bulletin de la Societe Francaise de Mineralogie, v56, 81-187 (1933) |
| 37300 | Schoellner, R., et al., Zeolites, v3, 149-154 (1983) |
| 37301 | Schoellner, R., et al., Zeolites, v3, 149-154 (1983) |
| 37302 | Schoellner, R., et al., Zeolites, v3, 149-154 (1983) |
| 37303 | Schoellner, R., et al., Zeolites, v3, 149-154 (1983) |
| 37304 | Schoellner, R., et al., Zeolites, v3, 149-154 (1983) |
| 37305 | Schoellner, R., et al., Zeolites, v3, 149-154 (1983) |
| 37306 | Schoellner, R., et al., Zeolites, v3, 149-154 (1983) |
| 37307 | Schoellner, R., et al., Zeolites, v3, 149-154 (1983) |
| 37308 | Schoellner, R., et al., Zeolites, v3, 149-154 (1983) |
| 37309 | Schoellner, R., et al., Zeolites, v3, 149-154 (1983) |
| 37449 | Bennett, J.M., et al., Materials Research Bulletin, v4, 7-14 (1969) |
| 38170 | Howell, P.A., Acta Crystallographica, v13, 737-741 (1960) |
| 38399 | Hambley, T.W., et al., Journal of Solid State Chemistry, v54, 1-9 (1984) |
| 38400 | Hambley, T.W., et al., Journal of Solid State Chemistry, v54, 1-9 (1984) |
| 38408 | Kvick, A., et al., Journal of Chemical Physics, v79, 2356-2362 (1983) |
| 39704 | Butikova, I.K., et al., Kristallografiya, v38, n4, 68-72 (1993) |
| 39705 | Butikova, I.K., et al., Kristallografiya, v38, n4, 68-72 (1993) |
| 39778 | Butikova, I.K., et al., Kristallografiya, v39, 426-429 (1994) |
| 39904 | Rastsvetaeva, R.K., Kristallografiya, v40, 812-815 (1995) |
| 40111 | Ross, C.R., et al., American Mineralogist, v75, 1249-1252 (1990) |
| 40126 | Vance, T.B.jr., et al., Journal of Physical Chemistry, v79, 2163-2167 (1975) |
| 40128 | Norby, P., et al., Acta Chemica Scandinavica, Series A:, v40, 500-506 (1986) |
| 40134 | Pluth, J.J., et al., Journal of Physical Chemistry, v83, 741-749 (1979) |
| 40136 | Fyfe, C.A., et al., Journal of the American Chemical Society, v111, 2470-2474 (1989) |
| 40137 | Fyfe, C.A., et al., Journal of Physical Chemistry, v94, 3718-3721 (1990) |
| 40138 | Mortier, W.J., et al., Journal of Physical Chemistry, v77, 2880-2885 (1973) |
| 40139 | Mortier, W.J., et al., Journal of Physical Chemistry, v77, 2880-2885 (1973) |
| 40140 | Mortier, W.J., et al., Journal of Physical Chemistry, v77, 2880-2885 (1973) |
| 40143 | Yang, P., et al., Journal of Solid State Chemistry, v123, 140-149 (1996) |
| 40409 | Kim, Y., et al., Journal of Physical Chemistry, v92, 5593-5596 (1988) |
| 40410 | Pluth, J.J., et al., Journal of Physical Chemistry, v93, 6516-6520 (1989) |
| 40421 | Pluth, J.J., et al., Journal of the American Chemical Society, v111, 1692-1698 (1989) |
| 40507 | Parise, J.B., et al., Journal of Physical Chemistry, v88, 1635-1640 (1984) |

| ICSD | Bibliographic reference |
| --- | --- |
| 40508 | Parise, J.B., et al., Journal of Physical Chemistry, v88, 1635-1640 (1984) |
| 40509 | Bergeret, G., et al., Journal of Physical Chemistry, v87, 1160-1165 (1983) |
| 40510 | Bergeret, G., et al., Journal of Physical Chemistry, v87, 1160-1165 (1983) |
| 40511 | Bergeret, G., et al., Journal of Physical Chemistry, v87, 1160-1165 (1983) |
| 40512 | Bergeret, G., et al., Journal of Physical Chemistry, v87, 1160-1165 (1983) |
| 40513 | Bergeret, G., et al., Journal of Physical Chemistry, v87, 1160-1165 (1983) |
| 40517 | Fyfe, C.A., et al., Journal of the American Chemical Society, v111, 2470-2474 (1989) |
| 40518 | Shepelev, Yu.F., et al., Kristallografiya, v33, 359-364 (1988) |
| 40532 | Gramlich-Meier, R., et al., Zeitschrift fuer Kristallographie, v169, 201-210 (1984) |
| 40533 | Ito, M., et al., Bulletin of the Chemical Society of Japan, v58, 3035-3036 (1985) |
| 40643 | Malinovskii, Yu.A., et al., Kristallografiya, v36, 571-576 (1991) |
| 40644 | Malinovskii, Yu.A., et al., Kristallografiya, v36, 571-576 (1991) |
| 40645 | Malinovskii, Yu.A., et al., Kristallografiya, v36, 577-583 (1991) |
| 40883 | Alberti, A., et al., Zeitschrift fuer Kristallographie, v178, 249-256 (1987) |
| 40885 | Bissert, G., et al., Zeitschrift fuer Kristallographie, v179, 357-371 (1987) |
| 40926 | Taylor, J.C., et al., Zeitschrift fuer Kristallographie, Kristallgeometrie, Kristallphysik, Kristallchemie, v176, 183-192 (1986) |
| 40927 | Fitch, A.N., et al., Journal of Physical Chemistry, v90, 1311-1318 (1986) |
| 40928 | Fitch, A.N., et al., Journal of Physical Chemistry, v90, 1311-1318 (1986) |
| 40929 | Fitch, A.N., et al., Journal of Physical Chemistry, v90, 1311-1318 (1986) |
| 40931 | Newsam, J.M., et al., Journal of Physical Chemistry, v90, 6858-6864 (1986) |
| 40932 | Newsam, J.M., et al., Journal of Physical Chemistry, v90, 6858-6864 (1986) |
| 40933 | Newsam, J.M., et al., Journal of Physical Chemistry, v90, 6858-6864 (1986) |
| 40934 | Heo, N.-H., et al., Journal of Physical Chemistry, v90, 3931-3935 (1986) |
| 40939 | Gies, H., Zeitschrift fuer Kristallographie, Kristallgeometrie, Kristallphysik, Kristallchemie, v175, 93-104 (1986) |
| 40940 | Alberti, A., et al., Zeitschrift fuer Kristallographie, Kristallgeometrie, Kristallphysik, Kristallchemie, v175, 249-256 (1986) |
| 40941 | Norby, P., et al., Chemistry of Materials, v12, 1473-1479 (2000) |
| 40942 | Mentzen, B.F., et al., Materials Research Bulletin, v22, 309-321 (1987) |
| 40943 | Mentzen, B.F., Materials Research Bulletin, v22, 337-343 (1987) |
| 41051 | Mentzen, B.F., Materials Research Bulletin, v27, 953-960 (1992) |
| 41052 | Mentzen, B.F., Materials Research Bulletin, v27, 953-960 (1992) |
| 41053 | Mentzen, B.F., Materials Research Bulletin, v27, 953-960 (1992) |
| 41054 | Mentzen, B.F., Materials Research Bulletin, v27, 953-960 (1992) |
| 41195 | Pechar, F., Crystal Research and Technology, v17, 1141-1144 (1982) |
| 41276 | Cabellal, R., et al., European Journal of Mineralogy, v5, 353-360 (1993) |
| 41394 | Jeanjean, J., et al., Journal of the Chemical Society. Faraday Transactions 1., v85, 2771-2783 (1989) |
| 41395 | Jeanjean, J., et al., Journal of the Chemical Society. Faraday Transactions 1., v85, 2771-2783 (1989) |
| 41396 | Jeanjean, J., et al., Journal of the Chemical Society. Faraday Transactions 1., v85, 2771-2783 (1989) |
| 41397 | Jeanjean, J., et al., Journal of the Chemical Society. Faraday Transactions 1., v85, 2771-2783 (1989) |
| 41398 | Jeanjean, J., et al., Journal of the Chemical Society. Faraday Transactions 1., v85, 2771-2783 (1989) |
| 41552 | McCusker, L.B., et al., Journal of Physical Chemistry, v85, 405-410 (1981) |
| 41553 | McCusker, L.B., et al., Journal of Physical Chemistry, v85, 405-410 (1981) |

| ICSD | Bibliographic reference |
|---|---|
| 55231 | Zanardi, S., et al., American Mineralogist, v89, 1033-1042 (2004) |
| 55232 | Zanardi, S., et al., American Mineralogist, v89, 1033-1042 (2004) |
| 55233 | Zanardi, S., et al., American Mineralogist, v89, 1033-1042 (2004) |
| 55234 | Zanardi, S., et al., American Mineralogist, v89, 1033-1042 (2004) |
| 55235 | Zanardi, S., et al., American Mineralogist, v89, 1033-1042 (2004) |
| 55236 | Zanardi, S., et al., American Mineralogist, v89, 1033-1042 (2004) |
| 55346 | Burton, A., et al., Chemistry - A European Journal, v9, 5737-5748 (2003) |
| 55347 | Burton, A., et al., Chemistry - A European Journal, v9, 5737-5748 (2003) |
| 55448 | Smolin, Yu.I., et al., Glass Physics and Chemistry, v29, 476-478 (2003) |
| 55474 | Colligan, M., et al., Journal of the American Chemical Society, v126, 12015-12022 (2004) |
| 55475 | Colligan, M., et al., Journal of the American Chemical Society, v126, 12015-12022 (2004) |
| 55476 | Colligan, M., et al., Journal of the American Chemical Society, v126, 12015-12022 (2004) |
| 55477 | Colligan, M., et al., Journal of the American Chemical Society, v126, 12015-12022 (2004) |
| 55478 | Colligan, M., et al., Journal of the American Chemical Society, v126, 12015-12022 (2004) |
| 55479 | Colligan, M., et al., Journal of the American Chemical Society, v126, 12015-12022 (2004) |
| 55480 | Colligan, M., et al., Journal of the American Chemical Society, v126, 12015-12022 (2004) |
| 55481 | Colligan, M., et al., Journal of the American Chemical Society, v126, 12015-12022 (2004) |
| 55482 | Colligan, M., et al., Journal of the American Chemical Society, v126, 12015-12022 (2004) |
| 55483 | Colligan, M., et al., Journal of the American Chemical Society, v126, 12015-12022 (2004) |
| 55484 | Colligan, M., et al., Journal of the American Chemical Society, v126, 12015-12022 (2004) |
| 55485 | Colligan, M., et al., Journal of the American Chemical Society, v126, 12015-12022 (2004) |
| 55486 | Colligan, M., et al., Journal of the American Chemical Society, v126, 12015-12022 (2004) |
| 55487 | Colligan, M., et al., Journal of the American Chemical Society, v126, 12015-12022 (2004) |
| 55488 | Colligan, M., et al., Journal of the American Chemical Society, v126, 12015-12022 (2004) |
| 55489 | Colligan, M., et al., Journal of the American Chemical Society, v126, 12015-12022 (2004) |
| 55490 | Colligan, M., et al., Journal of the American Chemical Society, v126, 12015-12022 (2004) |
| 55491 | Colligan, M., et al., Journal of the American Chemical Society, v126, 12015-12022 (2004) |
| 55910 | Celestian, A.J., et al., Chemistry of Materials, v16, 2244-2254 (2004) |
| 55948 | Dorset, D.L., et al., J. Phys. Chem. B, v108, 15216-15222 (2004) |
| 55949 | Kim Soo Yeon, et al., J. Phys. Chem. B, v107, 6938-6945 (2003) |
| 55950 | Kim Soo Yeon, et al., J. Phys. Chem. B, v107, 6938-6945 (2003) |
| 56301 | Lengauer, C.L., et al., Mineralogical Magazine, v61, 591-606 (1997) |
| 56320 | Koennecke, M., et al., Zeitschrift fuer Kristallographie, v201, 147-155 (1992) |
| 56321 | Koennecke, M., et al., Zeitschrift fuer Kristallographie, v201, 147-155 (1992) |
| 56478 | Meden, A., et al., Zeitschrift fuer Kristallographie, v212, 801-807 (1997) |
| 56479 | Meden, A., et al., Zeitschrift fuer Kristallographie, v212, 801-807 (1997) |
| 56657 | Finch, A.A., et al., Powder Diffraction, v10, n4, 243-247 (1995) |
| 56775 | Gies, H., et al., Zeitschrift fuer Kristallographie, v210, 475-480 (1995) |
| 56801 | Kaszkur, Z.A., et al., Journal of Physical Chemistry, v97, 426-431 (1993) |
| 56802 | Kaszkur, Z.A., et al., Journal of Physical Chemistry, v97, 426-431 (1993) |
| 57106 | Miehe, G., et al., Acta Crystallographica B, v49, 745-754 (1993) |
| 57126 | Gruenewald-Lueke, A., et al., Zeitschrift fuer Kristallographie - New Crystal Structures, v216, 655-656 (2001) |

| ICSD | Bibliographic reference |
| --- | --- |
| 57128 | Collet, P., et al., European Journal of Solid State Inorganic Chemistry, v28, 345-361 (1991) |
| 57132 | Marler, B., et al., Journal of Inclusion Phenomena, v4, 339-349 (1986) |
| 60187 | Kirfel, A., et al., Zeolites, v4, 140-146 (1984) |
| 60597 | Cheetham, A.K., et al., American Chemical Society: Symposium Series, v218, 132-142 (1983) |
| 60674 | Chao, K.-J., et al., Zeolites, v6, 35-38 (1986) |
| 60890 | Norby, P., et al., Acta Chemica Scandinavica, Series A:, v40, 500-506 (1986) |
| 60891 | Norby, P., et al., Acta Chemica Scandinavica, Series A:, v40, 500-506 (1986) |
| 60892 | Norby, P., et al., Acta Chemica Scandinavica, Series A:, v40, 500-506 (1986) |
| 61010 | Liu, Z., et al., Ranliao Huaxue Xuebao, v13, 106-113 (1985) |
| 61053 | Ito, M., et al., Acta Crystallographica C, v41, 1698-1700 (1985) |
| 61166 | Pluth, J.J., et al., Zeolites, v5, 74-80 (1985) |
| 61178 | Highcock, R.M., et al., Acta Crystallographica C, v41, 1391-1394 (1985) |
| 61187 | Gallezot, P., et al., Journal de Chimie et de Physique, v71, 155-163 (1974) |
| 61188 | Gallezot, P., et al., Journal de Chimie et de Physique, v71, 155-163 (1974) |
| 61189 | Gallezot, P., et al., Journal de Chimie et de Physique, v71, 155-163 (1974) |
| 61190 | Gallezot, P., et al., Journal de Chimie et de Physique, v71, 155-163 (1974) |
| 61242 | Artioli, G., et al., Acta Crystallographica C, v42, 937-942 (1986) |
| 61438 | Dimitrijevic, R., et al., Studies in Surface Science and Catalysis, v24, 453-458 (1985) |
| 61439 | Rinaldi, R., et al., Studies in Surface Science and Catalysis, v24, 481-492 (1985) |
| 61440 | Rinaldi, R., et al., Studies in Surface Science and Catalysis, v24, 481-492 (1985) |
| 61700 | Gallezot, P., et al., Journal of Catalysis, v26, 295-302 (1972) |
| 61701 | Gallezot, P., et al., Journal of Catalysis, v26, 295-302 (1972) |
| 62077 | Heo, N.-H., et al., Journal of the Chemical Society. Chemical Communications, v581, 1225-1226 (1987) |
| 62078 | Heo, N.-H., et al., Journal of the Chemical Society. Chemical Communications, v581, 1225-1226 (1987) |
| 62098 | Calligaris, M., et al., Zeolites, v6, 439-444 (1986) |
| 62274 | van Koningsveld, H., et al., Acta Crystallographica B, v43, 127-132 (1987) |
| 62293 | Mikheeva, M.G., et al., Kristallografiya, v31, 434-439 (1986) |
| 62315 | Vigdorchik, A.G., et al., Kristallografiya, v31, 879-882 (1986) |
| 62393 | Baur, W.H., et al., Zeitschrift fuer Kristallographie, v179, 281-304 (1987) |
| 62394 | Baur, W.H., et al., Zeitschrift fuer Kristallographie, v179, 281-304 (1987) |
| 62395 | Baur, W.H., et al., Zeitschrift fuer Kristallographie, v179, 281-304 (1987) |
| 62396 | Baur, W.H., et al., Zeitschrift fuer Kristallographie, v179, 281-304 (1987) |
| 62581 | Papiz, M.Z., et al., Acta Crystallographica C, v46, 172-173 (1990) |
| 62582 | LaPierre, R.B., et al., Zeolites, v5, 356-348 (1985) |
| 62599 | Calligaris, M., et al., Zeolites, v5, 317-319 (1985) |
| 62600 | Calligaris, M., et al., Zeolites, v5, 317-319 (1985) |
| 62691 | Fischer, R.X., et al., Journal of Physical Chemistry, v90, 4414-4423 (1986) |
| 62692 | Calligaris, M., et al., Zeolites, v6, 137-141 (1986) |
| 62693 | Calligaris, M., et al., Zeolites, v6, 137-141 (1986) |
| 62950 | Elsen, J., et al., Journal of Physical Chemistry, v91, 5800-5805 (1987) |
| 62951 | Elsen, J., et al., Journal of Physical Chemistry, v91, 5800-5805 (1987) |
| 62952 | Elsen, J., et al., Journal of Physical Chemistry, v91, 5800-5805 (1987) |

| ICSD | Bibliographic reference |
| --- | --- |
| 65498 | Barri, S.A.I., et al., Nature, v312, 533-534 (1984) |
| 65499 | Calestani, G., et al., Zeolites, v7, 59-62 (1987) |
| 65500 | Calestani, G., et al., Zeolites, v7, 54-58 (1987) |
| 65551 | Briscoe, N.A., et al., Zeolites, v8, 74-76 (1988) |
| 65624 | Smolin, Yu.I., et al., Acta Crystallographica B, v45, 124-128 (1989) |
| 65625 | Smolin, Yu.I., et al., Acta Crystallographica B, v45, 124-128 (1989) |
| 65626 | Smolin, Yu.I., et al., Acta Crystallographica B, v45, 124-128 (1989) |
| 65627 | Smolin, Yu.I., et al., Acta Crystallographica B, v45, 124-128 (1989) |
| 65628 | Smolin, Yu.I., et al., Acta Crystallographica B, v45, 124-128 (1989) |
| 65629 | Smolin, Yu.I., et al., Acta Crystallographica B, v45, 124-128 (1989) |
| 65671 | Fischer, R.X., et al., Acta Crystallographica C, v45, 983-989 (1989) |
| 65672 | Fischer, R.X., et al., Acta Crystallographica C, v45, 983-989 (1989) |
| 65788 | van Koningsveld, H., et al., Acta Crystallographica B, v45, 423-431 (1989) |
| 66041 | Xie, D., et al., Materials Research Society Symposia Proceedings, v111, 147-154 (1987) |
| 66063 | Corbin, D.R., et al., Journal of the American Chemical Society, v112, 4821-4830 (1990) |
| 66064 | Corbin, D.R., et al., Journal of the American Chemical Society, v112, 4821-4830 (1990) |
| 66065 | Corbin, D.R., et al., Journal of the American Chemical Society, v112, 4821-4830 (1990) |
| 66077 | Liu, X., et al., Cuihua Xuebao, v11, 196-203 (1990) |
| 66100 | Shepelev, Yu.F., et al., Zeolites, v10, 61-63 (1990) |
| 66101 | Shepelev, Yu.F., et al., Zeolites, v10, 61-63 (1990) |
| 66102 | Anderson, A.A., et al., Zeolites, v10, 32-37 (1990) |
| 66103 | Anderson, A.A., et al., Zeolites, v10, 32-37 (1990) |
| 66104 | Anderson, A.A., et al., Zeolites, v10, 32-37 (1990) |
| 66105 | Anderson, A.A., et al., Zeolites, v10, 32-37 (1990) |
| 66138 | van Dun, J.J., et al., Journal of Physics and Chemistry of Solids, v50, 469-477 (1989) |
| 66139 | van Dun, J.J., et al., Journal of Physics and Chemistry of Solids, v50, 469-477 (1989) |
| 66140 | van Dun, J.J., et al., Journal of Physics and Chemistry of Solids, v50, 469-477 (1989) |
| 66152 | Andries, K.J., et al., Zeolites, v11, 124-131 (1991) |
| 66153 | Krogh Andersen, I.G., et al., Zeolites, v11, 149-154 (1991) |
| 66156 | Norby, P., et al., Zeolites, v11, 248-253 (1991) |
| 66157 | Huddersman, K.D., et al., Zeolites, v11, 270-276 (1991) |
| 66158 | Shepelev, Yu.F., et al., Zeolites, v11, 287-292 (1991) |
| 66159 | Shepelev, Yu.F., et al., Zeolites, v11, 287-292 (1991) |
| 66160 | Shepelev, Yu.F., et al., Zeolites, v11, 287-292 (1991) |
| 66161 | Shepelev, Yu.F., et al., Zeolites, v11, 287-292 (1991) |
| 66162 | Shepelev, Yu.F., et al., Zeolites, v11, 287-292 (1991) |
| 66165 | Lin, J., et al., Zeolites, v11, 376-379 (1991) |
| 66166 | Lin, J., et al., Zeolites, v11, 376-379 (1991) |
| 66244 | Kim, Y., et al., Bulletin of the Korean Chemical Society, v9, 338-341 (1988) |
| 66245 | Kim, Y., et al., Bulletin of the Korean Chemical Society, v9, 338-341 (1988) |
| 66327 | Artioli, G., American Mineralogist, v77, 189-196 (1992) |
| 66328 | Artioli, G., American Mineralogist, v77, 189-196 (1992) |

| ICSD | Bibliographic reference |
| --- | --- |
| 66383 | Jang, S.B., et al., Journal of the Korean Chemical Society, v35, 630-635 (1991) |
| 66409 | Czjzek, M., et al., Journal of Physical Chemistry, v96, 1535-1540 (1992) |
| 66410 | Czjzek, M., et al., Journal of Physical Chemistry, v96, 1535-1540 (1992) |
| 66411 | Czjzek, M., et al., Journal of Physical Chemistry, v96, 1535-1540 (1992) |
| 66457 | Armbruster, T., et al., American Mineralogist, v76, 1872-1883 (1991) |
| 66458 | Armbruster, T., et al., American Mineralogist, v76, 1872-1883 (1991) |
| 66459 | Armbruster, T., et al., American Mineralogist, v76, 1872-1883 (1991) |
| 66460 | Armbruster, T., et al., American Mineralogist, v76, 1872-1883 (1991) |
| 66461 | Armbruster, T., et al., American Mineralogist, v76, 1872-1883 (1991) |
| 66475 | Lievens, J.L., et al., Journal of Physics and Chemistry of Solids, v53, 1163-1169 (1992) |
| 66476 | Lievens, J.L., et al., Journal of Physics and Chemistry of Solids, v53, 1163-1169 (1992) |
| 66477 | Lievens, J.L., et al., Journal of Physics and Chemistry of Solids, v53, 1163-1169 (1992) |
| 66478 | Lievens, J.L., et al., Journal of Physics and Chemistry of Solids, v53, 1163-1169 (1992) |
| 66479 | Lievens, J.L., et al., Journal of Physics and Chemistry of Solids, v53, 1163-1169 (1992) |
| 66648 | Li, X., et al., Fenzi Cuihua, v6, n2, 104-111 (1992) |
| 66687 | Akizuki, M., et al., European Journal of Mineralogy, v5, 839-843 (1993) |
| 66689 | Stahl, K., et al., European Journal of Mineralogy, v5, 851-856 (1993) |
| 66742 | Kato, M., et al., X-sen Bunseki no Shinpo, v25, 111-120 (1993) |
| 66743 | Kato, M., et al., X-sen Bunseki no Shinpo, v25, 111-120 (1993) |
| 66844 | Lobo, R.F., et al., Chemistry of Materials, v8, 2409-2411 (1996) |
| 66857 | Schroepfer, L., et al., European Journal of Mineralogy, v9, 53-65 (1997) |
| 66858 | Schroepfer, L., et al., European Journal of Mineralogy, v9, 53-65 (1997) |
| 66859 | Schroepfer, L., et al., European Journal of Mineralogy, v9, 53-65 (1997) |
| 67005 | Kim, Y., et al., Bulletin of the Korean Chemical Society, v10, n4, 349-352 (1989) |
| 67006 | Kim, Y., et al., Bulletin of the Korean Chemical Society, v10, n4, 349-352 (1989) |
| 67007 | Pickering, I.J., et al., Journal of Catalysis, v119, 261-265 (1989) |
| 67029 | Newsam, J.M., Journal of Physical Chemistry, v93, 7689-7694 (1989) |
| 67030 | Newsam, J.M., Journal of Physical Chemistry, v93, 7689-7694 (1989) |
| 67031 | Newsam, J.M., Journal of Physical Chemistry, v93, 7689-7694 (1989) |
| 67095 | Butikova, I.K., et al., Kristallografiya, v34, 1136-1140 (1989) |
| 67096 | Butikova, I.K., et al., Kristallografiya, v34, 1136-1140 (1989) |
| 67097 | Butikova, I.K., et al., Kristallografiya, v34, 1136-1140 (1989) |
| 67098 | Butikova, I.K., et al., Kristallografiya, v34, 1141-1145 (1989) |
| 67099 | Butikova, I.K., et al., Kristallografiya, v34, 1141-1145 (1989) |
| 67104 | Shepelev, Yu.F., et al., Kristallografiya, v34, 1302-1304 (1989) |
| 67210 | Pechar, F., Zeitschrift fuer Kristallographie, v189, 191-194 (1989) |
| 67330 | Song, S.H., et al., Journal of the Korean Chemical Society, v33, n5, 452-458 (1989) |
| 67331 | Song, S.H., et al., Journal of the Korean Chemical Society, v33, n5, 452-458 (1989) |
| 67465 | McCusker, L.B., Journal of Applied Crystallography, v21, 305-310 (1988) |
| 67592 | Song, S.H., et al., Journal of Physical Chemistry, v96, 10937-10941 (1992) |
| 67621 | Song, S.H., et al., Journal of Physical Chemistry, v96, 10937-10941 (1992) |
| 67622 | Song, S.H., et al., Journal of Physical Chemistry, v96, 10937-10941 (1992) |

| ICSD | Bibliographic reference |
| --- | --- |
| 84464 | Yeom, Y.H., et al., Journal of Physical Chemistry, v101, 6914-6920 (1997) |
| 84465 | Shibata, W., et al., Journal of Physical Chemistry, v101, 9022-9026 (1997) |
| 84466 | Jang, S.B., et al., Journal of Physical Chemistry, v101, 9041-9045 (1997) |
| 84467 | Jang, S.B., et al., Journal of Physical Chemistry, v101, 9041-9045 (1997) |
| 84468 | Anderson, P.A., et al., Journal of Physical Chemistry, v10, 9892-9900 (1997) |
| 84469 | Anderson, P.A., et al., Journal of Physical Chemistry, v10, 9892-9900 (1997) |
| 84470 | Anderson, P.A., et al., Journal of Physical Chemistry, v10, 9892-9900 (1997) |
| 84471 | Anderson, P.A., et al., Journal of Physical Chemistry, v10, 9892-9900 (1997) |
| 84532 | Pang, W.-Q., et al., Gaodeng Xuexiao Huaxue Xuebao, v5, n3, 375-380 (1984) |
| 84580 | Weidenthaler, C., et al., Acta Crystallographica B, v53, 429-439 (1997) |
| 84581 | Weidenthaler, C., et al., Acta Crystallographica B, v53, 429-439 (1997) |
| 84582 | Weidenthaler, C., et al., Acta Crystallographica B, v53, 429-439 (1997) |
| 84583 | Weidenthaler, C., et al., Acta Crystallographica B, v53, 429-439 (1997) |
| 84584 | Weidenthaler, C., et al., Acta Crystallographica B, v53, 429-439 (1997) |
| 84585 | Weidenthaler, C., et al., Acta Crystallographica B, v53, 429-439 (1997) |
| 84586 | Weidenthaler, C., et al., Acta Crystallographica B, v53, 429-439 (1997) |
| 84587 | Weidenthaler, C., et al., Acta Crystallographica B, v53, 440-443 (1997) |
| 84588 | Weidenthaler, C., et al., Acta Crystallographica B, v53, 440-443 (1997) |
| 84589 | Weidenthaler, C., et al., Acta Crystallographica B, v53, 440-443 (1997) |
| 84590 | Weidenthaler, C., et al., Acta Crystallographica B, v53, 444-450 (1997) |
| 84591 | Weidenthaler, C., et al., Acta Crystallographica B, v53, 444-450 (1997) |
| 84592 | Weidenthaler, C., et al., Acta Crystallographica B, v53, 444-450 (1997) |
| 84593 | Weidenthaler, C., et al., Acta Crystallographica B, v53, 444-450 (1997) |
| 85119 | van Koningsveld, H., et al., Microporous Materials, v9, 71-81 (1997) |
| 85176 | Stolz, J., et al., Neues Jahrbuch fuer Mineralogie. Monatshefte, v1997, n3, 131-144 (1997) |
| 85443 | Kim, Y., et al., Zeolites, v18, 325-333 (1997) |
| 85447 | Alberti, A., et al., Zeolites, v19, 349-352 (1997) |
| 85448 | Vezzalini, G., et al., Zeolites, v19, 75-79 (1997) |
| 85454 | Pechar, F., et al., Crystal Research and Technology, v21, n8, 1029-1034 (1986) |
| 85468 | Lobo, R.F., et al., Journal of the American Chemical Society, v119, 3732-3744 (1997) |
| 85469 | Lobo, R.F., et al., Journal of the American Chemical Society, v119, 3732-3744 (1997) |
| 85470 | Lobo, R.F., et al., Journal of the American Chemical Society, v119, 3732-3744 (1997) |
| 85471 | Lobo, R.F., et al., Journal of the American Chemical Society, v119, 3732-3744 (1997) |
| 85472 | Lobo, R.F., et al., Journal of the American Chemical Society, v119, 3732-3744 (1997) |
| 85473 | Lobo, R.F., et al., Journal of the American Chemical Society, v119, 3732-3744 (1997) |
| 85474 | Lobo, R.F., et al., Journal of the American Chemical Society, v119, 3732-3744 (1997) |
| 85475 | Lobo, R.F., et al., Journal of the American Chemical Society, v119, 3732-3744 (1997) |
| 85511 | Artioli, G., et al., Zeolites, v6, 361-366 (1986) |
| 85514 | Fischer, R.X., et al., Zeolites, v6, 378-387 (1986) |
| 85515 | Fischer, R.X., et al., Zeolites, v6, 378-387 (1986) |
| 85517 | Eddy, M.M., et al., Zeolites, v6, 449-454 (1986) |
| 85543 | Gualtieri, A., et al., American Mineralogist, v83, 590-606 (1998) |

| ICSD | Bibliographic reference |
|---|---|
| 85544 | Gualtieri, A., et al., American Mineralogist, v83, 590-606 (1998) |
| 85545 | Gualtieri, A., et al., American Mineralogist, v83, 590-606 (1998) |
| 85546 | Gualtieri, A., et al., American Mineralogist, v83, 590-606 (1998) |
| 85547 | Gualtieri, A., et al., American Mineralogist, v83, 590-606 (1998) |
| 85548 | Gualtieri, A., et al., American Mineralogist, v83, 590-606 (1998) |
| 85549 | Gualtieri, A., et al., American Mineralogist, v83, 590-606 (1998) |
| 85550 | Effenberger, H., et al., American Mineralogist, v83, 607-617 (1998) |
| 85568 | Lee, S.H., et al., Bulletin of the Korean Chemical Society, v19, n1, 98-103 (1998) |
| 85580 | Chen, C.-Y., et al., Chemistry - A European Journal, v4, n7, 1312-1323 (1998) |
| 85586 | Diaz-Cabanas, M.J., et al., Chemical Communications, v1998, 1881-1882 (1998) |
| 85612 | Feuerstein, M., et al., Chemistry of Materials, v10, 2197-2204 (1998) |
| 85613 | Feuerstein, M., et al., Chemistry of Materials, v10, 2197-2204 (1998) |
| 85614 | Feuerstein, M., et al., Chemistry of Materials, v10, 2197-2204 (1998) |
| 85615 | Feuerstein, M., et al., Chemistry of Materials, v10, 2197-2204 (1998) |
| 85620 | Lee, Y.-J., et al., Chemistry of Materials, v10, 2561-2570 (1998) |
| 85621 | Lee, Y.-J., et al., Chemistry of Materials, v10, 2561-2570 (1998) |
| 85622 | Lee, Y.-J., et al., Chemistry of Materials, v10, 2561-2570 (1998) |
| 85623 | Lee, Y.-J., et al., Chemistry of Materials, v10, 2561-2570 (1998) |
| 85624 | Lee, Y.-J., et al., Chemistry of Materials, v10, 2561-2570 (1998) |
| 85625 | Lee, Y.-J., et al., Chemistry of Materials, v10, 2561-2570 (1998) |
| 85696 | Yang, P., et al., European Journal of Mineralogy, v10, 461-471 (1998) |
| 86203 | Heo, N.-H., et al., J. Phys. Chem. B, v102, 17-23 (1998) |
| 86204 | Haniffa, R.M., et al., J. Phys. Chem. B, v102, 2688-2695 (1998) |
| 86205 | Haniffa, R.M., et al., J. Phys. Chem. B, v102, 2688-2695 (1998) |
| 86206 | Camblor, M.A., et al., J. Phys. Chem. B, v102, 44-51 (1998) |
| 86279 | van Koningsveld, H., et al., Zeolites, v10, 235-242 (1990) |
| 86508 | Jirak, Z., et al., Journal of Physics and Chemistry of Solids, v41, 1089-1095 (1980) |
| 86509 | Jirak, Z., et al., Journal of Physics and Chemistry of Solids, v41, 1089-1095 (1980) |
| 86510 | Jirak, Z., et al., Journal of Physics and Chemistry of Solids, v41, 1089-1095 (1980) |
| 86511 | Jirak, Z., et al., Journal of Physics and Chemistry of Solids, v41, 1089-1095 (1980) |
| 86547 | Marler, B., et al., Zeolites, v15, 388-399 (1995) |
| 86548 | Marler, B., et al., Zeolites, v15, 388-399 (1995) |
| 86549 | Marler, B., et al., Zeolites, v15, 388-399 (1995) |
| 86621 | Kim, Y., et al., Bulletin of the Korean Chemical Society, v19, n11, 1222-1227 (1998) |
| 86633 | Porcher, F., et al., C.R. Acad. Sci. Paris, T. 1, Serie II, v1, 701-708 (1998) |
| 86634 | Porcher, F., et al., C.R. Acad. Sci. Paris, T. 1, Serie II, v1, 701-708 (1998) |
| 86642 | Ikeda, T., et al., Chemistry of Materials, v10, 3996-4004 (1998) |
| 86643 | Ikeda, T., et al., Chemistry of Materials, v10, 3996-4004 (1998) |
| 86644 | Ikeda, T., et al., Chemistry of Materials, v10, 3996-4004 (1998) |
| 86687 | Bu Xian-Hui, et al., Journal of the American Chemical Society, v120, 13389-13397 (1998) |
| 86688 | Bu Xian-Hui, et al., Journal of the American Chemical Society, v120, 13389-13397 (1998) |
| 86741 | Barrett, P.A., et al., Journal of Materials Chemistry, v8, n10, 2263-2268 (1998) |

| ICSD | Bibliographic reference |
| --- | --- |
| 90725 | Olson, D.H., et al., J. Phys. Chem. B, v104, 4844-4848 (2000) |
| 90727 | Heo, N.-H., et al., J. Phys. Chem. B, v104, 8372-8381 (2000) |
| 90728 | Heo, N.-H., et al., J. Phys. Chem. B, v104, 8372-8381 (2000) |
| 90729 | Zhu Lin, et al., J. Phys. Chem. B, v105, 12221-12221 (2001) |
| 91530 | Smolin, Yu.I., et al., Kristallografiya, v45, n1, 27-31 (2000) |
| 91531 | Smolin, Yu.I., et al., Kristallografiya, v45, n1, 27-31 (2000) |
| 91662 | Tripathi, A., et al., Microporous and Mesoporous Materials, v34, 273-279 (2000) |
| 91663 | Lee, Y.-J., et al., Microporous and Mesoporous Materials, v34, 255-271 (2000) |
| 91664 | Lee, Y.-J., et al., Microporous and Mesoporous Materials, v34, 255-271 (2000) |
| 91665 | Lee, Y.-J., et al., Microporous and Mesoporous Materials, v34, 255-271 (2000) |
| 91667 | Healey, A.M., et al., Microporous and Mesoporous Materials, v37, 153-163 (2000) |
| 91668 | Healey, A.M., et al., Microporous and Mesoporous Materials, v37, 165-174 (2000) |
| 91669 | Stolz, J., et al., Microporous and Mesoporous Materials, v37, 233-242 (2000) |
| 91671 | Klap, G.J., et al., Microporous and Mesoporous Materials, v38, 403-412 (2000) |
| 91672 | Klap, G.J., et al., Microporous and Mesoporous Materials, v38, 403-412 (2000) |
| 91673 | Klap, G.J., et al., Microporous and Mesoporous Materials, v38, 403-412 (2000) |
| 91674 | Klap, G.J., et al., Microporous and Mesoporous Materials, v38, 403-412 (2000) |
| 91675 | Zhu, L., et al., Microporous and Mesoporous Materials, v39, 187-193 (2000) |
| 91676 | Zhu, L., et al., Microporous and Mesoporous Materials, v39, 187-193 (2000) |
| 91678 | Fyfe, C.A., et al., Microporous and Mesoporous Materials, v39, 291-305 (2000) |
| 91679 | Kirchner, R.M., et al., Microporous and Mesoporous Materials, v39, 319-332 (2000) |
| 91680 | Kirchner, R.M., et al., Microporous and Mesoporous Materials, v39, 319-332 (2000) |
| 91681 | Kirchner, R.M., et al., Microporous and Mesoporous Materials, v39, 319-332 (2000) |
| 91684 | Dalconi, M.C., et al., Microporous and Mesoporous Materials, v39, 423-430 (2000) |
| 91685 | Dalconi, M.C., et al., Microporous and Mesoporous Materials, v39, 423-430 (2000) |
| 91689 | Bae, D.H., et al., Microporous and Mesoporous Materials, v40, 219-232 (2000) |
| 91690 | Bae, D.H., et al., Microporous and Mesoporous Materials, v40, 232-245 (2000) |
| 91691 | Bae, D.H., et al., Microporous and Mesoporous Materials, v40, 232-245 (2000) |
| 91692 | Choi, E.Y., et al., Microporous and Mesoporous Materials, v40, 247-255 (2000) |
| 91693 | Marra, G.L., et al., Microporous and Mesoporous Materials, v40, 85-94 (2000) |
| 91694 | Marra, G.L., et al., Microporous and Mesoporous Materials, v40, 85-94 (2000) |
| 91695 | Sacerdoti, M., et al., Microporous and Mesoporous Materials, v41, 107-118 (2000) |
| 91696 | Sacerdoti, M., et al., Microporous and Mesoporous Materials, v41, 107-118 (2000) |
| 91697 | Sacerdoti, M., et al., Microporous and Mesoporous Materials, v41, 107-118 (2000) |
| 91698 | Sacerdoti, M., et al., Microporous and Mesoporous Materials, v41, 107-118 (2000) |
| 91699 | Sacerdoti, M., et al., Microporous and Mesoporous Materials, v41, 107-118 (2000) |
| 91700 | Sacerdoti, M., et al., Microporous and Mesoporous Materials, v41, 107-118 (2000) |
| 91702 | Nery, J.G., et al., Microporous and Mesoporous Materials, v41, 281-293 (2000) |
| 91703 | Nery, J.G., et al., Microporous and Mesoporous Materials, v41, 281-293 (2000) |
| 91704 | Nery, J.G., et al., Microporous and Mesoporous Materials, v41, 281-293 (2000) |
| 91705 | Nery, J.G., et al., Microporous and Mesoporous Materials, v41, 281-293 (2000) |
| 91706 | Lee, S.H., et al., Microporous and Mesoporous Materials, v41, 49-59 (2000) |

| ICSD | Bibliographic reference |
|---|---|
| 200026 | Firor, R.L., et al., Journal of the American Chemical Society, v99, 1112-1117 (1977) |
| 200027 | Firor, R.L., et al., Journal of the American Chemical Society, v99, 1112-1117 (1977) |
| 200148 | Firor, R.L., et al., Journal of the American Chemical Society, v99, 6249-6253 (1977) |
| 200149 | Kim, Y., et al., Journal of the American Chemical Society, v99, 7055-7057 (1977) |
| 200150 | Kim, Y., et al., Journal of the American Chemical Society, v99, 7055-7057 (1977) |
| 200151 | Kim, Y., et al., Journal of the American Chemical Society, v99, 7057-7059 (1977) |
| 200152 | Firor, R.L., et al., Journal of the American Chemical Society, v99, 7059-7061 (1977) |
| 200253 | Subramanian, V., et al., Journal of Physical Chemistry, v81, 2249-2251 (1977) |
| 200272 | Firor, R.L., et al., Journal of the American Chemical Society, v100, 976-978 (1978) |
| 200273 | Firor, R.L., et al., Journal of the American Chemical Society, v100, 978-980 (1978) |
| 200274 | Firor, R.L., et al., Journal of the American Chemical Society, v100, 3091-3096 (1978) |
| 200275 | Firor, R.L., et al., Journal of the American Chemical Society, v100, 3091-3096 (1978) |
| 200276 | Kim, Y., et al., Journal of the American Chemical Society, v100, 3801-3805 (1978) |
| 200277 | McCusker, L.B., et al., Journal of the American Chemical Society, v100, 5052-5057 (1978) |
| 200278 | McCusker, L.B., et al., Journal of the American Chemical Society, v100, 5052-5057 (1978) |
| 200279 | Kim, Y., et al., Journal of the American Chemical Society, v100, 6989-6997 (1978) |
| 200280 | Kim, Y., et al., Journal of the American Chemical Society, v100, 6989-6997 (1978) |
| 200281 | Kim, Y., et al., Journal of the American Chemical Society, v100, 6989-6997 (1978) |
| 200282 | Kim, Y., et al., Journal of the American Chemical Society, v100, 6989-6997 (1978) |
| 200283 | Kim, Y., et al., Journal of the American Chemical Society, v100, 6989-6997 (1978) |
| 200284 | Kim, Y., et al., Journal of the American Chemical Society, v100, 6989-6997 (1978) |
| 200361 | Subramanian, V., et al., Journal of the American Chemical Society, v100, 2911-2913 (1978) |
| 200362 | Subramanian, V., et al., Journal of the American Chemical Society, v100, 2911-2913 (1978) |
| 200456 | Firor, R.L., et al., Journal of the American Chemical Society, v99, 4039-4044 (1977) |
| 200457 | Firor, R.L., et al., Journal of the American Chemical Society, v99, 4039-4044 (1977) |
| 200506 | Shepelev, Yu.F., et al., Kristallografiya, v24, 469-474 (1979) |
| 200507 | Shepelev, Yu.F., et al., Kristallografiya, v24, 469-474 (1979) |
| 200521 | Smolin, Yu.I., et al., Kristallografiya, v24, 461-468 (1979) |
| 200522 | Smolin, Yu.I., et al., Kristallografiya, v24, 461-468 (1979) |
| 200575 | Kim, Y., et al., Journal of Physical Chemistry, v82, 1071-1077 (1978) |
| 200576 | Kim, Y., et al., Journal of Physical Chemistry, v82, 1071-1077 (1978) |
| 200578 | Firor, R.L., et al., Journal of Physical Chemistry, v82, 1650-1655 (1978) |
| 200579 | Firor, R.L., et al., Journal of Physical Chemistry, v82, 1650-1655 (1978) |
| 200581 | Pluth, J.J., et al., Journal of Physical Chemistry, v83, 741-749 (1979) |
| 200582 | Pluth, J.J., et al., Journal of Physical Chemistry, v83, 741-749 (1979) |
| 200585 | Subramanian, V., et al., Journal of Physical Chemistry, v83, 2166-2169 (1979) |
| 200634 | Pearce, J.R., et al., Journal of the Chemical Society. Faraday Transactions 1., v75, 898-906 (1979) |
| 200635 | Pearce, J.R., et al., Journal of the Chemical Society. Faraday Transactions 1., v75, 898-906 (1979) |
| 200636 | Pearce, J.R., et al., Journal of the Chemical Society. Faraday Transactions 1., v75, 898-906 (1979) |
| 200951 | Alberti, A., et al., Physics and Chemistry of Minerals (Germany), v2, 365-375 (1978) |
| 201050 | Subramanian, V., et al., Journal of Physical Chemistry, v84, 2928-2933 (1980) |
| 201051 | Subramanian, V., et al., Journal of Physical Chemistry, v84, 2928-2933 (1980) |

| ICSD | Bibliographic reference |
|---|---|
| 280761 | Kongshaug, K.O., et al., Journal of Materials Chemistry, v11, 1242-1247 (2001) |
| 280762 | Kongshaug, K.O., et al., Journal of Materials Chemistry, v11, 1242-1247 (2001) |
| 281217 | Bull, I., et al., Journal of the American Chemical Society, v125, 4342-4349 (2003) |
| 281218 | Bull, I., et al., Journal of the American Chemical Society, v125, 4342-4349 (2003) |
| 281219 | Bull, I., et al., Journal of the American Chemical Society, v125, 4342-4349 (2003) |
| 281220 | Bull, I., et al., Journal of the American Chemical Society, v125, 4342-4349 (2003) |
| 281221 | Bull, I., et al., Journal of the American Chemical Society, v125, 4342-4349 (2003) |
| 281222 | Bull, I., et al., Journal of the American Chemical Society, v125, 4342-4349 (2003) |
| 281223 | Bull, I., et al., Journal of the American Chemical Society, v125, 4342-4349 (2003) |
| 281224 | Bull, I., et al., Journal of the American Chemical Society, v125, 4342-4349 (2003) |
| 281225 | Bull, I., et al., Journal of the American Chemical Society, v125, 4342-4349 (2003) |
| 281226 | Bull, I., et al., Journal of the American Chemical Society, v125, 4342-4349 (2003) |
| 281227 | Bull, I., et al., Journal of the American Chemical Society, v125, 4342-4349 (2003) |
| 281228 | Bull, I., et al., Journal of the American Chemical Society, v125, 4342-4349 (2003) |
| 281229 | Bull, I., et al., Journal of the American Chemical Society, v125, 4342-4349 (2003) |
| 281537 | Corma, A., et al., Angew. Chem. Int. ed., v42, n10, 1156-1159 (2003) |
| 281735 | Zheng Nan-Feng, et al., Science, v298, 2366-2369 (2002) |
| 281746 | Zheng Nan-Feng, et al., Science, v298, 2366-2369 (2002) |
| 281750 | Zheng Nan-Feng, et al., Science, v298, 2366-2369 (2002) |
| 281751 | Zheng Nan-Feng, et al., Science, v298, 2366-2369 (2002) |
| 281752 | Zheng Nan-Feng, et al., Science, v298, 2366-2369 (2002) |
| 281753 | Zheng Nan-Feng, et al., Science, v298, 2366-2369 (2002) |
| 281754 | Zheng Nan-Feng, et al., Science, v298, 2366-2369 (2002) |
| 281757 | Zheng Nan-Feng, et al., Science, v298, 2366-2369 (2002) |
| 281758 | Zheng Nan-Feng, et al., Science, v298, 2366-2369 (2002) |
| 281759 | Zheng Nan-Feng, et al., Science, v298, 2366-2369 (2002) |
| 410595 | Wagner, P., et al., Angew. Chem. Int. ed., v38, 1269-1272 (1999) |
| 410596 | Wagner, P., et al., Angew. Chem. Int. ed., v38, 1269-1272 (1999) |
| 411155 | Milanesio, M., et al., J. Phys. Chem. B, v104, n43, 9951-9953 (2000) |
| 413853 | Zanardi, S., et al., Angew. Chem. Int. ed., v43, n37, 4933-4937 (2004) |